%% file: hyperfine_h3p.tex
\documentclass[aps,jcp,superscriptaddress,amsmath,amssymb,preprint]{revtex4-2}

\usepackage{graphicx}
\usepackage{dcolumn}
\usepackage{bm}

\usepackage{rotating}
\newcolumntype{.}{D{.}{.}{8}}

\usepackage[utf8]{inputenc}
\usepackage{physics}
\usepackage{amsmath, amssymb}
\usepackage{tabularx,booktabs,array,dcolumn}
\usepackage{setspace}
\usepackage{vmargin}
\usepackage{xcolor}
\usepackage{graphicx,psfrag,subfigure}

\usepackage{float}
\usepackage[all]{xy}

\usepackage{mathtools}
\usepackage{physics}
\usepackage[english]{babel}
\usepackage{bm}

\usepackage[utf8]{inputenc}
\usepackage{physics}
\usepackage{amsmath, amssymb}
\usepackage{tabularx,booktabs,array,dcolumn}
\usepackage{setspace}
\usepackage{vmargin}
\usepackage{xcolor}
\usepackage{graphicx,psfrag,subfigure}

\usepackage{url}
\usepackage{float}
\usepackage[all]{xy}
\usepackage{multirow}
\usepackage{comment}

\usepackage[round-mode=places,round-precision=4]{siunitx} 
\usepackage[version=4]{mhchem}

\usepackage{physics}
\usepackage{url}

\newcommand{\bos}[1]{\boldsymbol{#1}}

\newcommand{\iim}{\mathrm{i}}
\newcommand{\cm}{cm$^{-1}$}

\def\tT{\text{T}}
\def\ZARE{\text{Z}}

\def\nnuc{N} 
\def\munuc{\mu_\text{N}}

\usepackage[unicode]{hyperref}
\hypersetup{
   unicode=true,          
   plainpages=false,
   colorlinks=true,       
   citecolor=blue,        
}

\def\ZARE{{\mathrm{Z}}}

\def\MF{{\mathrm{MF}}}

\def\dd{{\mathrm{d}}}
\def\detg{\tilde{g}}
\def\ncm{\text{NCM}}
\def\Hrv{H_\text{rv}}

\def\srot{\mathcal{M}} 
\def\ss{\mathcal{R}} 
\def\dss{\mathcal{D}} 
\def\gmol{\mathcal{G}} 
\def\shield{\sigma} 

\def\omone{\alpha}
\def\omtwo{\beta}
\def\omthree{\gamma}
\def\bq{\boldsymbol{q}}

\def\ext{\text{ext}}
\def\mB{\mathcal{B}}
\def\Sf{I}
\def\hS{\hat{I}}
\def\barT{\bar{T}}
\def\MF{}

\def\rone{r_{12}}
\def\rtwo{r_{13}}
\def\tangle{\theta_{213}}

\def\Atwop{A_2'}

\def\Ep{E'}
\def\Epp{E''}

\def\tK{{K}} 
\def\wK{{\tilde{K}}} 

\usepackage[unicode]{hyperref}
\usepackage{soul}
\hypersetup{
   unicode=true,          
   plainpages=false,
   colorlinks=true,       
   linkcolor=black,          
   linkcolor=blue,          
   citecolor=blue,        
   urlcolor=blue           
}

\begin{document}

\title{Hyperfine rovibrational states of H$_3^+$ in a weak external magnetic field}

\author{Gustavo Avila}
\affiliation{ELTE, E\"otv\"os Lor\'and University, Institute of Chemistry, %
P\'azm\'any P\'eter s\'et\'any 1/A, %
1117 Budapest, Hungary}

\author{Ayaki Sunaga}%
\affiliation{ELTE, E\"otv\"os Lor\'and University, Institute of Chemistry, %
P\'azm\'any P\'eter s\'et\'any 1/A, %
1117 Budapest, Hungary}

\author{Stanislav Komorovsky}%
\affiliation{%
Institute of Inorganic Chemistry, Slovak Academy of Sciences, Dubravska cesta 9, 84536 Bratislava, Slovakia
}%

\author{Edit M\'atyus}%
\email{edit.matyus@ttk.elte.hu}
\affiliation{ELTE, E\"otv\"os Lor\'and University, Institute of Chemistry, %
P\'azm\'any P\'eter s\'et\'any 1/A, %
1117 Budapest, Hungary}

\date{\today}

\begin{abstract}
\noindent 
Rovibrational energies, wave functions, and Raman transition moments are reported for the lowest-energy states of the H$_3^+$ molecular ion including the magnetic couplings of the proton spins and molecular rotation in the presence of a weak external magnetic field.
The rovibrational-hyperfine-Zeeman Hamiltonian matrix is constructed and diagonalized using the rovibrational eigenstates and the proton spin functions.
The developed methodology can be used to compute hyperfine-Zeeman effects also for higher-energy rovibrational excitations of H$_3^+$ and other polyatomic molecules. 
These developments will guide future experiments extending quantum logic spectroscopy to polyatomic systems.
\end{abstract}

\maketitle

\clearpage
%
%
Recent developments in quantum technology operating on a molecular platform \cite{SiMeNaHeWi20,NaMeWi20,LiLeLeCh20,NaMeSiWi20,SiWi23},
have made it possible to cool and coherently manipulate single diatomic ions \cite{ChKuHuPlLeLe17,SiMeNaHeWi20,NaMeWi20,LiLeLeCh20,NaMeSiWi20,SiWi23,ScHoStMiSeHoKi23,singleH2p}. In Ref.~\cite{ChKuHuPlLeLe17}, the cooling and selective quantum state preparation of the $^{40}$CaH$^+$
`spectroscopy ion' was carried out through its entangled (quantized) translational motion within the ion trap with the $^{40}$Ca$^+$, `logic ion'. As a result, the rovibrational hyperfine-Zeeman level structure of $^{40}$CaH$^+$ was probed by stimulated Raman spectroscopy with unprecedented control and precision.
Similar spectroscopic experiments using quantum logic \cite{ScRoLaItBeWi05} have already been successfully realized for $^{24}$MgH$^+$~\cite{WoWaHeGeShSc16}, 
CaH$^+$~\cite{ChKuHuPlLeLe17,ChCoKuLiHaPlFoDiLeLe20}, N$_2^+$~\cite{SiMeNaHeWi20}, and most recently, H$_2^+$~\cite{singleH2p}. In principle, quantum logic spectroscopy may be extended to polyatomic systems. 
In this work, we develop a computational approach for modelling rovibrational-hyperfine-Zeeman effects in polyatomic systems to guide future experiments. 
We report the first numerical applications for the simplest polyatomic cation, H$_3^+$.

The quantum mechanical motion of molecular systems has been described within the non-relativistic and Born-Openheimer (BO) approximations. These two approximations define two major fields 
(a) electronic structure theory, which deals with the numerical solution of the electronic Schrödinger equation;
and (b) quantum nuclear motion theory, which is about the numerical solution of the rovibrational Schrödinger equation on a potential energy surface (PES).
For predictions and guidance for spectroscopy, astrophysics, and potentially, quantum technology, 
it is useful to think about the list of rovibrational energies (wave functions) and transition moments as a `quantum dynamics database' \cite{Te16,OwYa18,CsFaSz21,YaYaZaYuKu22,OwZaChYuTeYa20}. Then, this database, after including the necessary couplings, allows us to simulate further interactions over the relevant rovibrational subspace. 
The numerical solution of the (field-free) rovibrational Schrödinger equation of H$_3^+$ has a long history \cite{TeSu84,TeSu85,TeSu86,BrCa93,BrTrCaCo94}.
In this work, we go beyond the standard field-free approach by accounting for the interaction between molecular rotation, nuclear spins, and an external magnetic field, \emph{i.e.,} we consider the nuclear spin-rotation tensor, the rotational g-tensor, the nuclear spin-spin coupling tensor, and the NMR shielding tensor in conjunction with the quantum mechanical treatment of molecular rotations and vibrations.

Regarding earlier work on hyperfine modelling, without any external magnetic field, the spin-rotation coupling was computed for rovibrational transitions of H$_2$D$^+$ and HD$_2^+$ \cite{JePaSpSa97}. The possibility of observing ortho-para transitions in the water molecule was also studied by including the spin-spin and spin-rotation coupling in the variational rovibrational computations \cite{MiTe04,YaYaZaYuKu22}. Recent work reported modelling hyperfine effects in diatomic systems of astrophysical relevance \cite{QuYuTe22}.

%
%
\vspace{0.5cm}
The rovibrational Schrödinger equation, 
$
  \Hrv \psi_n = E_n \psi_n 
$
can be efficiently solved by replacing the laboratory-fixed (LF) Cartesian coordinates, 
$\bos{R}_i=(R_{iX},R_{iY},R_{iZ})$, $(i=1,\ldots,N)$ of the nuclei with physically motivated internal, $q_1,\ldots,q_{3N-6}$, orientational, $\Omega=(\omone,\omtwo,\omthree)$, and overall translational coordinates corresponding to the nuclear centre of mass, 
$\bos{R}_\ncm = \sum_{i=1}^{N} \frac{m_i}{m_{1\ldots N}}  \bos{R}_i$
with the $m_i$ masses associated to the nuclei and $m_{1\ldots N} = \sum_{i=1}^N m_i$.
The rotational coordinates, $\Omega=(\alpha,\beta,\gamma)$, define the orientation of the 
the body-fixed (BF) frame through the relation \cite{Za98} 
\begin{align}
  \bos{R}_i
  =
  \bos{O}(\Omega) \bos{r}_i + \bos{R}_\ncm \; ,\quad i = 1,\ldots,N \;,
  \label{eq:frame}
\end{align}
where the origin of the $\bos{r}_i$ body-fixed Cartesian coordinates is at the centre of mass.
In what follows, capital letters $A,B=X,Y,Z$ are used for LF directions, and small letters, $a,b=x,y,z$ label BF directions.
The $q_1,\ldots,q_{3N-6}$ internal coordinates are defined as scalar (and vector) products of the $\bos{r}_i$ body-fixed Cartesian coordinates~\cite{Su02}.

The rovibrational Hamiltonian is the sum of the kinetic energy operator, now written in the curvilinear Podolsky form \cite{MaCzCs09}, and the $V$ potential energy depending on the internal coordinates,
\begin{align}
  \Hrv
  =
  &\frac{1}{2} 
  \sum_{k=1}^{3N-3} \sum_{l=1}^{3N-3} 
    \detg^{-1/4} \hat{p}_k G_{kl} \detg^{1/2} \hat{p}_l \detg^{-1/4} 
  + V \; .
  \label{eq:Hrv}    
\end{align}
$\hat{p}_k=-\iim\partial / \partial q_k\ (k=1,\ldots,3N-6)$ is the momentum (in atomic units) conjugate to the $q_k$ curvilinear internal coordinate, 
$\hat{p}_{3N-6+a}=\hat{J}_a\ (a=1,2,3)$ labels the $a$th component of the rotational angular momentum in the body-fixed frame.
$\bos{G}=\bos{g}^{-1}$ and $\detg=\text{det}\bos{g}$ include the $\bos{g}$ mass-weighted metric tensor corresponding to the curvilinear coordinate transformation.
The Hamiltonian in Eq.~\eqref{eq:Hrv} is understood with the simple $\dd V=\dd q_1 \ldots \dd q_{3N-6}$ volume element for calculating integrals (the Jacobian is merged in the Hamiltonian and in the wave function). Ref.~\cite{MaDaAv23} reviews further details of the formalism and the (quasi)-variational solution technique of the rovibrational Schrödinger equation as implemented in the GENIUSH program \cite{MaCzCs09,FaMaCs11} used in this work.

The rovibrational eigenfunctions are also eigenfunctions of the $\hat{J}^2$, $\hat{J}_Z$ rotational angular momentum operators with quantum numbers $J$ and $M$, and the space inversion operator.
We compute the rovibrational wave function as a linear combination of products of vibrational basis functions, $f_v(\bos{q})$, and the $|J\tK M\rangle_\Omega$ symmetric-top functions,
\begin{align}
  \Psi_n^{(JM)}(\Omega,\bq)
  =
  \sum_{v \tK}
    c_{n,v \tK}^{(JM)}
    f_v(\bq)
    |J \tK M\rangle_\Omega
  \label{eq:rovibwf}
\end{align}
with $\tK,M=-J,-J+1\ldots,J-1,J$ for all $J$ (the Wang combinations \cite{Wa29} are discussed in \cite{som}),
where 
\begin{align}
  |J \tK M\rangle_\Omega
  =
  \bar{D}^{\text{Z},J\ast}_{M \tK}(\alpha,\beta,\gamma) 
  =
  \sqrt{\frac{2J+1}{8\pi^2}} D^{\text{Z},J\ast}_{M \tK}(\alpha,\beta,\gamma) \; ,
  \label{eq:WignerDZ}
\end{align}
is a normalized Wigner $D$ function (the normalization is indicated by the bar). We use the same phase conventions as Zare \cite{Za98}, hence the Z superscript.

%
%
\vspace{0.5cm}
The interaction of the molecular rotation and the nuclear spin magnetic moments within the system and with the external magnetic field, $\mB^\ext$, is described by the hyperfine-Zeeman Hamiltonian  terms, 
\begin{align}
  H_\text{hf} 
  &= 
  -\sum_{i=1}^{\nnuc}
  \sum_{A,B=1}^3
    \hat{I}_{iA} \srot_{iA,B} \hat{J}_B   
      +
  \sum_{i=1}^{\nnuc}  
  \sum_{j>i}^{\nnuc}
  \sum_{A,B=1}^3
    \hat{I}_{iA} (\dss_{iA,jB}+\ss_{iA,jB}) \hat{I}_{jB} \; ,
  \label{eq:Hhf}
  \\    
  H_\text{Z}   
  &=
  -
  \sum_{A,B=1}^3  
    \munuc
      \hat{J}_{A} \gmol_{AB} \mB^\ext_B      
  -
  \sum_{i=1}^{\nnuc}
  \sum_{A,B=1}^3
    g_i\munuc
    \hat{I}_{iA} (\delta_{A,B}-\shield_{iA,B}) \mB^\ext_B   
    \; ,
  \label{eq:HZ}
\end{align}
which is parameterized in this work through the contributions leading order in $1/c$ \cite{Ra53,HeJaRu99,HeJoOl,Fl74}. 
Furthermore, in the general case, the interaction of the nuclear quadrupole moment with the electric field gradient also contributes to the hyperfine structure, but it is not relevant for the present application for H$_3^+$, since the proton has a zero quadrupole moment.
Furthermore, only weak external fields are considered in this work, so higher-order magnetic effects are assumed to be negligible.

In Eqs.~\eqref{eq:Hhf} and \eqref{eq:HZ},
$\srot_{iA,B}$, $\ss_{iA,jB}$, $\gmol_{AB}$, and $\shield_{iA,B}$ are the LF elements of the nuclear spin-rotation, the indirect nuclear spin-spin coupling, the rotational g-tensor, and the magnetic shielding tensor, respectively. 
$\dss_{iA,jB}$ is the direct spin-spin coupling, 
\begin{align}
  \dss_{iA,jB}
  =
  -\frac{g_i g_j \munuc^2}{c^2}
  \left\lbrace%
     \frac{1}{R_{ij}^3}
     \left[%
       \frac{3 {R}_{ij,A} {R}_{ij,B}}{R_{ij}^2}
       -\delta_{A,B}
     \right]
     +\frac{8\pi}{3}\delta(\bos{R}_{ij}) 
  \right\rbrace \; 
  \label{eq:dss}
\end{align}
with the $\bos{R}_{ij}$ displacement vector of nuclei $i$ and $j$ in the LF frame and $R_{ij}=|\bos{R}_{ij}|$. The contact term proportional to $\delta(\bos{R}_{ij})$ is indicated but not used in the computations, since it is expected to give a negligible contribution ($<10^{-10}$~\cm) to the hyperfine splittings. (This contact term is not directly calculable in a BO-based description, because the PES is infinite at the coalescence of two protons. 
For H$_2^+$, $\langle \delta{(\bos{R}_{12})}\rangle$ is available from three-body computations \cite{Fr99}.)

The listed tensors account for couplings including the nuclear (here: proton) spin angular momenta, $\hat{\bos{I}}_i$, and the molecular rotational angular momentum, $\hat{\bos{J}}$.
$\munuc=1/(2m_\text{p})$ (in SI-based atomic units) is the nuclear magneton, $c$ is the speed of light, $m_\text{p}$ is the proton mass, and throughout this work, $g_i=g_\text{p}$ is the proton g-factor.

We aim to compute the eigenstates of the $H_\text{rv}+H_\text{hf}+H_\text{Z}$ rovibrational-hyperfine-Zeeman Hamiltonian.
Since the hyperfine terms, Eq.~\eqref{eq:Hhf}, couple the rotational ($\hat{\bos{J}}$) and the proton spin angular momenta ($\hat{\bos{I}}=\hat{\bos{I}}_1+\hat{\bos{I}}_2+\hat{\bos{I}}_3$), the $\hat{\bos{F}}^2$ and $\hat{F}_Z$ total angular momenta, $\hat{\bos{F}}=\hat{\bos{J}}+\hat{\bos{I}}$, are the conserved quantities with the quantum numbers $F$ and $M_F$.
If a non-zero $\bos{\mB}^\ext$ external magnetic field is present, which we assume to be oriented along the $Z$ LF axis, the Zeeman terms, Eq.~\eqref{eq:HZ}, couple different $F$ states, only $\hat{F}_Z$ is conserved, and $M_F$ is the exact quantum number of the theory.
We also note that for an external magnetic field, there is no space-inversion symmetry, therefore different parity states can mix.

To guide possible experimental detection of the hyperfine-Zeeman-rovibrational splittings, we compute Raman transition moments as 
\begin{align}
  \mathcal{R} = \sum_{A,B=X,Y,Z} |\langle\Psi_\text{f}|\alpha_{AB}|\Psi_\text{i}\rangle|^2 \; ,
  \label{eq:polar}
\end{align}
where $\alpha_{AB}$ is the $(A,B)$th LF component of the electric dipole polarizability operator. The hyperfine-Zeeman levels are not degenerate for the $M$ magnetic quantum number ($\hat{J}_Z$), nor for the $M_I$ spin quantum number ($\hat{I}_Z$). So, nuclear spin weights cannot be used, and the transition moments are explicitly computed for all pairs of the rovibrational-hyperfine-Zeeman wave functions.

%
%
\vspace{0.5cm}
First, we numerically solve the rovibrational Schrödinger equation, and then, construct the matrix representation of the rovibrational-hyperfine-Zeeman Hamiltonian. So, we need to compute the matrix elements connecting the products of the (magnetic-coupling-free) rovibrational eigenfunctions and the proton spin functions.
The LF magnetic coupling matrices and the dipole polarizability matrix depend both on the orientational and the internal coordinates.
To construct the matrix representation,
we consider a general `coupling matrix' $T$ in the LF frame. $T$ can be expressed with the BF-frame coupling matrix, $\bos{t}(\bos{q})$, which is computed from the molecular properties and structure, and the $\bos{O}(\Omega)$ orthogonal rotation matrix, which connects the LF and BF frames, Eq.~\eqref{eq:frame}, 
\begin{align}
  \bos{T}(\Omega,\bq) 
  =
  \bos{O}(\Omega)
  \bos{t}(\bq)
  \bos{O}^\tT(\Omega) \; .
\end{align}
By exploiting this relation, we can express the orientational dependence of $T$ using the orthonormal Wigner $D$-matrices, which is written for a general (non-symmetric) second-rank tensor as
\begin{align}
  T_{AB}(\Omega,\bq) 
  =
  \sum_{l=0}^2
    \sum_{p=-l}^l
    \sum_{q=-l}^l
      \bar{D}^{\text{Z},l}_{pq}(\Omega) 
      \bar{T}^{AB}_{lpq}(\bq)
\end{align}
for the $A,B=X,Y,Z$ LF directions. So, the angular dependence is carried by the Wigner $D$ matrices, while $\bar{T}^{AB}_{lpq}(\bq)$ is a linear combination of 
the body-fixed $t_{ij}(\bq)$ $(i,j=x,y,z)$ elements, which depend only on the $\bos{q}$ internal coordinates.
Then, we compute the matrix elements of $T$ with the rovibrational eigenfunctions, Eq.~\eqref{eq:rovibwf}, as
\begin{align}
  &\langle 
    \psi_{n'}^{(J'M')}
    | T_{AB} | 
    \psi_{n}^{(JM)}
  \rangle 
  =
  \sum_{\tK'=-J'}^{J'}
  \sum_{\tK=-J}^{J}
  \sum_{v',v}
    c^{\ast(J'M')}_{n',v'\tK'} c^{(JM)}_{n,v\tK} 
    \nonumber \\
  &\quad\quad\quad
    \sum_{l=0}^2
      \sum_{p=-l}^l
      \sum_{q=-l}^l
        \langle J'\tK'M' | \bar{D}^{\text{Z},l}_{pq}(\Omega) | J\tK M\rangle_{\Omega} \ 
        \langle f_{v'}(\bq) | \bar{T}_{lpq}^{AB}(\bq) | f_v(\bq)\rangle_{\bq} \; .
  \label{eq:int1}
\end{align}
If the hyperfine-Zeeman coupling term contains an angular momentum operator factor, then, we have
\begin{align}
  &\langle 
    \psi_{n'}^{(J'M')}
    | \hat{J}_A T_{AB} | 
    \psi_{n}^{(JM)}
  \rangle 
  =
  \sum_{\tK'=-J'}^{J'}
  \sum_{\tK=-J}^J
    c^{\ast(J'M')}_{n',v'\tK'} c^{(JM)}_{n,v\tK} 
    \nonumber \\
  &\quad\quad\quad
    \sum_{l=0}^2
      \sum_{p=-l}^l
      \sum_{q=-l}^l
        \langle J' \tK' M' | \hat{J}_A \bar{D}^{\text{Z},l}_{pq}(\Omega) | J \tK M\rangle_{\Omega} \ 
        \langle %
          f_{v'}(\bq)
          | \bar{T}_{lpq}^{AB}(\bq) | 
          f_{v}(\bq)
        \rangle_{\bq} \; ,
  \label{eq:int2}
\end{align}
where we act with $\hat{J}_A$ on the bra and proceed with the evaluation of the sum. 
Alternatively, in a $T_{AB} \hat{J}_B$-type term, we act with $\hat{J}_B$ on the ket. 
Finally, we calculate the $\Omega$ integral using the known analytic integral expression of the product of three Wigner $D$ functions  (\emph{e.g.,} Eq.~3.118 of Ref.~\cite{Za98}) resulting in 3$j$ symbols.
The integration for the internal coordinate dependence is straightforwardly computed 
with the quadrature used for the vibrational part of the problem.
Regarding spin integration, the matrices for the spin-1/2 angular momentum components correspond to the LF frame directions and are calculated with the usual definition of the spin operators.

%
%
\begin{figure}
  \includegraphics[width=5cm]{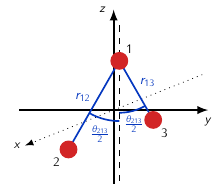} \\
  \caption{%
    The internal coordinates, $r_{12},r_{13},$ and $\theta_{213}$ and the $x,y,z$ body-fixed frame axes (bisector embedding) used in the rovibrational computations of H$_3^+$. The protons are fixed in the $yz$ plane, and the origin of the frame is at the nuclear centre of mass. The body-fixed $z$ axis is parallel with the bisector (dashed line) of the 2-1-3 proton angle. 
    \label{fig:coords}
  }
\end{figure}

\vspace{0.5cm}
Regarding the computational details, the rovibrational Schrödinger equation of H$_3^+$ was solved using the GENIUSH program \cite{MaCzCs09,FaMaCs11} and the GLH3P PES \cite{pes12} based on variationally optimized explicitly correlated Gaussian functions. 
We use the proton mass throughout this work, and 
we only mention that the empirical model of using different proton masses for rotation and vibration \cite{PoTe99} had been found to provide rovibrational transition energies closer to the experimental values, which was explained by (empirically) modelling small non-adiabatic effects \cite{Ku07,FuSzMaFaCs13}, which can be derived by perturbation theory \cite{MaTe19,FeMa19,FeKoMa20,MaFe22}.
We use the values of the physical constants according to the CODATA~2022 recommendation~\cite{codata22}.

During the rovibrational computations, we used the $(\rone,\rtwo,\cos\tangle)$ coordinates as internal coordinates and the bisector embedding as the BF frame (Fig.~\ref{fig:coords}). 
To construct the matrix representation of the (ro)vibrational Hamiltonian, the $\rone$ and $\rtwo$ degrees of freedom were described with a potential-optimized (PO) discrete variable representation (DVR) \cite{WeCa92,EcCl92}
based on the $L^{\alpha}_n$ associated Laguerre polynomials with $\alpha=2$. We used Jacobi DVR, based on the $J^{\alpha,\beta}_n$ Jacobi polynomials with $\alpha=\beta=0.05$,  for the $\cos\tangle$ degree of freedom. 
We converged all rotational energies up to $J=3$ corresponding to the ground vibrational state to  $10^{-6}$~\cm\ using the direct product grid including (61,61,91) points,  which was necessary to have an unbiased assessment of all hyperfine and magnetic contributions in later steps.

For the magnetic properties \cite{HeJaRu99,CaPuHaGa09,PuStGa10}, 
we used the CFOUR \cite{cfourshort,cfour-jcp20} (and the DALTON \cite{daltonshort,dalton_paper}) electronic structure program package(s) to compute the spin-rotation matrix \cite{GaRuHe96,GaSu97} and the rotational $g$ matrix \cite{GaRuKa07}. The nuclear magnetic shielding (NMR shielding) \cite{GaSt95} and the indirect spin-spin coupling matrices were also determined \cite{HeJaRu99,AuGa01}. In exploratory computations, they were found to give negligibly small contributions to the rovibrational-hyperfine-Zeeman levels ($<10^{-10}$~\cm), and hence they were neglected from the further computations. 
The final magnetic properties were obtained
with the coupled-cluster singles and doubles (CCSD) approximation \cite{StGaWaBa91}
and gauge-invariant atomic orbitals \cite{Lo37,HeJaRu99}.
Based on convergence tests, we used the cc-pVTZ basis set for the production runs and computed all couplings over the entire quadrature grid  ($61\times61\times91 = 338\ 611$ points) of the vibrational part. 
In this direct product grid, convergence issues were encountered during the CCSD computation for some highly distorted geometries. In these cases, the magnetic tensors were set to zero, because the rovibrational wave functions have a very small amplitude at these extreme structures, so these points would give a negligibly small contribution to the matrix elements. 
For the evaluation of the rovibrational-hyperfine-Zeeman transition matrix elements, 
all magnetic coupling matrices computed with the electronic structure packages were rotated to the body-fixed frame (bisector embedding, Fig.~\ref{fig:coords}) used in the rovibrational computations. 

To construct the rovibrational-hyperfine-Zeeman Hamiltonian matrix,
we used the product of the rovibrational eigenstates, including all rotational states up to $J^\text{max}=3$ corresponding to the vibrational ground state and the eight proton spin functions.
This basis set included $(1+3\times3+5\times 5 +7\times 7)\times 8= 2\ 288$ proton-spin-rovibrational functions ($\tK,M=-J,-J+1,\ldots,J-1,J)$, which is also the size of the Hamiltonian matrix. 
To ensure strict hermiticity of the Hamiltonian matrix (the hyperfine-Zeeman terms are not in a manifestly hermitian form), the matrix representation of the spin-rotation term was computed for 
\begin{align}
  -\sum_{i=1}^{\nnuc}
  \sum_{A,B=1}^3
    \hat{I}_{iA} \srot_{iA,B} \hat{J}_B     
  =
  -\frac{1}{2}
  \sum_{i=1}^{\nnuc}
  \sum_{A,B=1}^3
    \left[%
      \hat{I}_{iA} \srot_{iA,B} \hat{J}_B
      +
      \hat{J}_B \srot^\tT_{B,iA} \hat{I}_{iA} 
    \right]
\end{align}
similarly to Ref.~\cite{YaYaZaYuKu22}. All other terms were used as in Eqs.~\eqref{eq:Hhf} and \eqref{eq:HZ} and the Hamiltonian matrix fulfilled the hermiticity condition to machine precision ($10^{-14}$). The hyperfine-Zeeman-rotational eigenvalues and eigenvectors were computed by direct diagonalization of this matrix.

\vspace{0.5cm}At the end of the computation, we selected the Pauli-allowed rovibrational-hyperfine-Zeeman states. 
Since the protons are spin-1/2 fermions, the physically relevant, rovibrational-spin wave functions transform according to the $A_2$ irreducible representation (irrep) of the $S_3$ permutation group (the subgroup of the $D_{3\text{h}}(\text{M})$ molecular symmetry group \cite{BuJe98} excluding space inversion).
So, we retained only those singly degenerate rovibrational-hyperfine-Zeeman states for which the expectation value of the `$P_{23}$' permutation was $-1$. For the present coordinate system (Fig.~\ref{fig:coords}), $P_{23}\rone=\rtwo$, $P_{23}\rtwo=\rone$, and $P_{23}\tangle=\tangle$, and the transformation rules for the body-fixed frame and the symmetric-top eigenfunctions are also simple \cite{som}. 
In future work, we plan to use the Pekeris coordinates \cite{Pe58,WeCa94}, which are linear combinations of the proton-proton distances (with independent intervals), and hence, a direct product grid using identical 1D grids is closed under all $S_3$ operations, so a full $A_2$ symmetry projector can be straightforwardly defined over the DVR grid. %
Alternatively, we will transform the (ro)vibrational wave function from (not PO-)DVR to the finite basis representation (FBR) and perform all necessary symmetry operations in FBR.

\vspace{0.5cm}
The dipole polarizability matrix was computed at the CCSD/aug-cc-pV5Z level of theory using the DALTON program. The matrix printed by DALTON was rotated to the bisector embedding (Fig.~\ref{fig:coords}) for every nuclear geometry.
The computation was repeated at every point of a direct product grid of $17 \times 17\times 17 = 4\ 913 $ nuclear configurations incremented with some additional, randomly generated geometries. 
After discarding problematic points (close to linear configurations and large bond distances) 
$6\ 958$ geometries were used to fit an analytic function, which 
represented the polarizability matrix with a relative error less than $0.1\%$ \cite{som}. This representation is expected to provide a good prediction of the hyperfine-Zeeman transitions observable with Raman spectroscopy.

\begin{figure}
  \hspace{-0.5cm} 
  \scalebox{1.10}{%
    \includegraphics[scale=1.]{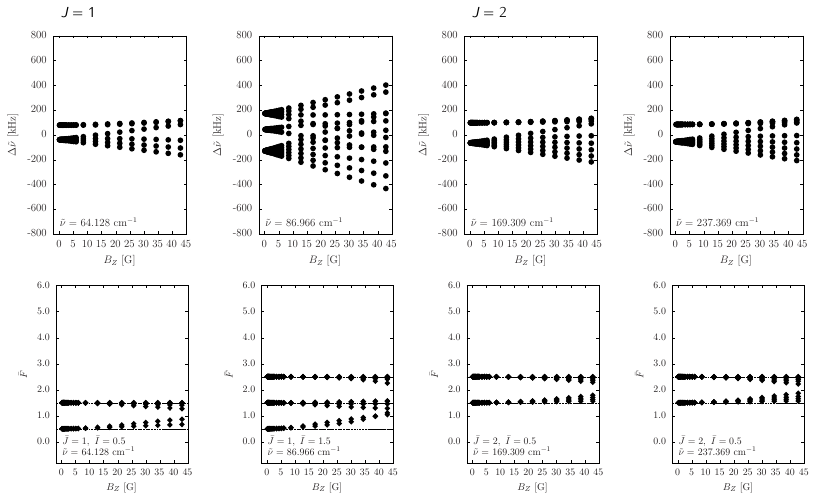}
  }
\caption{%
  Hyperfine-Zeeman-rotational levels of H$_3^+$ corresponding to the $J=1$ and $J=2$ rotational states. 
  The $\langle \hat{\bos{F}^2} \rangle=\bar{F}(\bar{F}+1)$ expectation value was computed for the actual state, and $\bar{F}$ is plotted in the lower panels. 
  The external magnetic field is along the $Z$ laboratory-fixed axis, $\mathcal{\bos{B}}^\text{ext}=(0,0,B_Z)$.
  \label{fig:states12}
}
\end{figure}

%
%
\begin{figure}
  \hspace{-0.5cm} 
  \scalebox{1.10}{%
    \includegraphics[scale=1.]{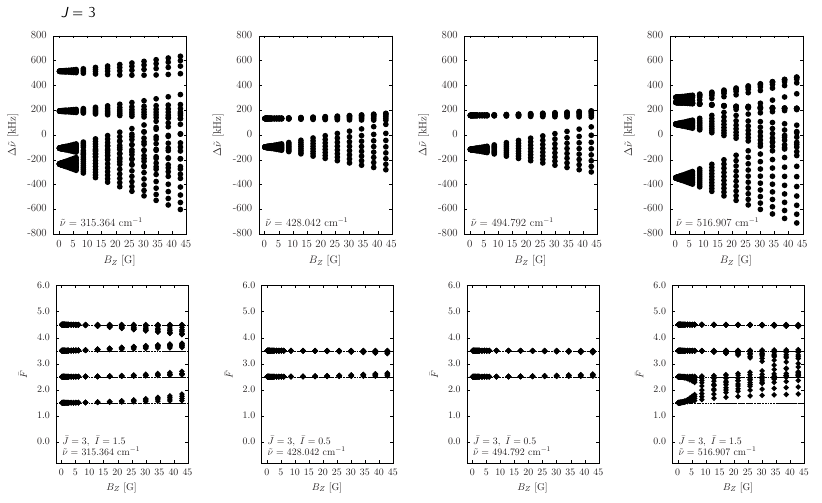}
  }
\caption{%
  Hyperfine-Zeeman-rotational levels of H$_3^+$ corresponding to the $J=3$ rotational states. 
  The $\langle \hat{\bos{F}}^2 \rangle=\bar{F}(\bar{F}+1)$ expectation value was computed for the actual state, and $\bar{F}$ is plotted in the lower panels.
  The external magnetic field is along the $Z$ laboratory-fixed axis, $\mathcal{\bos{B}}^\text{ext}=(0,0,B_Z)$.
  \label{fig:states3}
}
\end{figure}

\vspace{0.5cm}
The hyperfine-magnetic energies were obtained by direct diagonalization of the rovibrational-hyperfine-Zeeman Hamiltonian for a series of external magnetic field values, between 0 and 45~G  (0 and 4.5 mT), and the external magnetic field was oriented along the $Z$ LF axis.
The computed hyperfine-Zeeman energies are visualized in Figures~\ref{fig:states12} and \ref{fig:states3}.
The coupling between states of different rotational and total proton spin angular momentum quantum numbers, $J$ and $I$, is small, so, $J$ and $I$ continue to be very good (approximate) quantum numbers. We observe stronger mixing for different $F$ states at higher magnetic field strengths as shown in the lower panels of the figures, \emph{e.g.,} for the hyperfine-Zeeman manifold of the rotational state at $\tilde\nu=86.966$~\cm.

\begin{figure}
  \scalebox{0.89}{%
    \includegraphics[scale=1.]{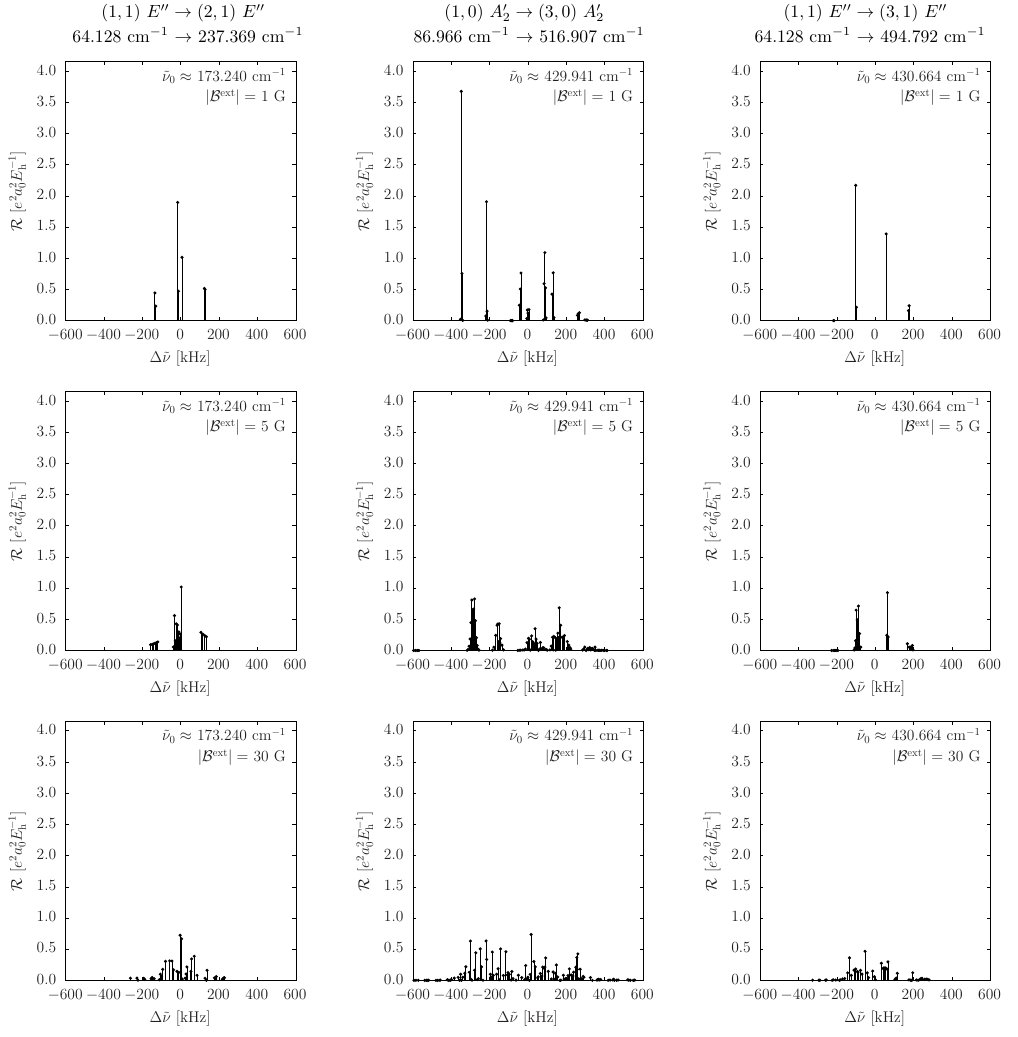}
  }
\caption{%
  Raman transition moments connecting the hyperfine-Zeeman-rotational levels of H$_3^+$ for an external magnetic field oriented along the $Z$ laboratory-fixed axis, $\mathcal{\bos{B}}^\text{ext}=(0,0,B_Z)$ with $B_Z=1,5,$ and 30~G.
  The lower and upper state labels and rotational energies \cite{som} are also shown at the top of the figure.
  \label{fig:Raman}
}
\end{figure}

Figure~\ref{fig:Raman} presents the predicted Raman transition moments within the hyperfine-Zeeman manifolds of %
the $(1,1)\ E''  \rightarrow (2,1)\ E''$  transition at 173.240~\cm, 
the $(1,0)\ A_2' \rightarrow (3,0)\ A_2'$ transition at 429.941~\cm, and
the $(1,1)\ E''  \rightarrow (3,1)\ E''$ transition at 430.792~\cm. 
At lower fields, $|\mB^\ext|=1$~G (0.1~mT), a simple pattern is seen, dominated by the spin-rotation and direct spin-spin coupling. At higher fields, the transitions are split by the direct spin-magnetic and rotation-magnetic couplings. 
For $\mB^\ext=30$~G (3~mT),  we see a congested spectrum indicating multiple competing effects.
The spin-rotation and rotation-magnetic couplings become more important for higher $J$ values.

\begin{figure}
  \scalebox{0.95}{%
    \includegraphics[scale=1.]{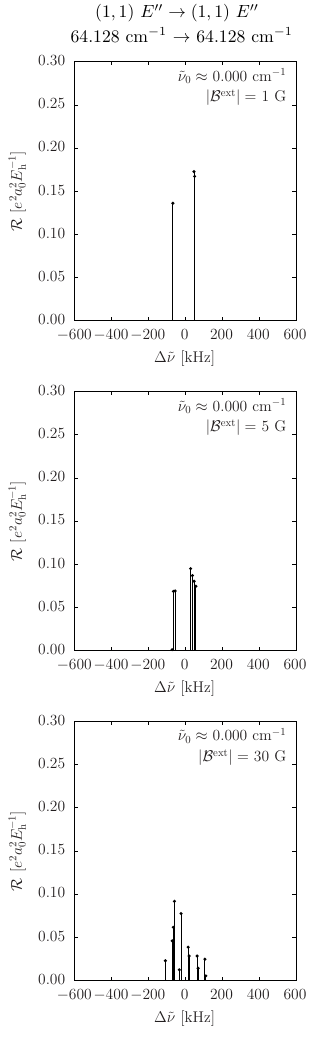}
  }
\caption{%
  Raman transition moments within the hyperfine-Zeeman manifold of the rovibrational ground state, $(1,1)\ E''$ (64.128~\cm), of H$_3^+$ for an external magnetic field oriented along the $Z$ laboratory-fixed axis, $\mathcal{\bos{B}}^\text{ext}=(0,0,B_Z)$ with $B_Z=1,5,$ and 30~G.
  \label{fig:Ramangs}
}
\end{figure}
We also note that non-vanishing Raman transition moments are seen also within the hyperfine-Zeeman components within the same rotational band. As an example, Fig.~\ref{fig:Ramangs} shows the transition moments connecting the hyperfine-Zeeman states within the $(1,1)\ E''$ (64.128~\cm) ground-state rotational manifold.

\vspace{0.5cm}
The present work reported the development of a new computational methodology to account for hyperfine-Zeeman effects in the high-resolution rovibrational spectrum of polyatomic molecules. Numerical results were presented for the simplest polyatomic system, H$_3^+$, using a variational rovibrational approach and an accurate \emph{ab initio} potential energy surface. The hyperfine-magnetic coupling matrices were computed by standard electronic structure theory including the leading-order contributions in $1/c$. The magnetic property computations were carried out over the vibrational grid representation to avoid fitting errors. 
It will be interesting to compare the computed results with future experiments. It is expected that the current results are sufficiently accurate to guide the experimental design and detection, which may be the first extension of quantum logic spectroscopy to a polyatomic system.
If necessary, the present level of numerical error control and convergence of the electronic energy values underlying the potential energy surface can be further improved, and further corrections, including the leading-order regularized relativistic~\cite{RaFeMaMa24} and quantum electrodynamics corrections with precise evaluation of the Bethe logarithm~\cite{FeMa23} can be included. 
The precision of the current PES (based on variational explicitly correlated Gaussian computations \cite{pes12}) can be further improved.
Furthermore, the hyperfine-magnetic coupling matrices could also be computed using an explicitly correlated Gaussian basis representation.
As to the rovibrational part, the hyperfine-Zeeman splittings can be computed for a large number of rovibrational states, since the integration of the property tensors with the large (gigantic) DVR grid used in this work is very accurate.
Furthermore, the hyperfine-Zeeman Hamiltonian matrix,
for the magnetic field values used in this work,
can be efficiently diagonalized in smaller matrix blocks without losing accuracy. In this respect, it will be interesting to study the interaction of energetically close-lying (or even near degenerate) rovibrational states with different $J$ values through the hyperfine--Zeeman couplings.

The H$_3^+$ molecular ion is of particular interest as the simplest polyatomic system. The computed results will be used to guide the experiments and also to test molecular quantum mechanics beyond the simplest systems, \emph{i.e.,} H$_2^+$ and isotopologues \cite{KoHiKa06,KaHaHiZhKo20,HaKoHiKa22}. In turn, H$_3^+$, as a polyatomic species, provides orders of magnitude higher complexity due to its rich rovibrational internal structure \cite{BoMiZoPoSaPoPeYuTe23}, and thus, potentially high-quality data to measure and a greater variety of (rovibrational) properties to control.

\vspace{0.5cm}
\begin{acknowledgments}
\noindent Financial support of the European Research Council through a Starting Grant (No.~851421) is gratefully acknowledged.
A.S. thanks the European Union’s Horizon 2022 research and innovation programme under the Marie Skłodowska-Curie Grant Agreement No.~101105452.
We also thank the financial support of the Hungarian National Research, Development, and Innovation Office (FK~142869), and the Slovak Grant Agency APVV (Grant No. APVV-22-0488 to S.K.).
\end{acknowledgments}

%

\input{hyperfine_h3p_som_inc}

\end{document}

%% file: hyperfine_h3p_som_inc.tex
\clearpage
\begin{center}
{\large
\textbf{Supplemental Material }
}\\[0.25cm]
{\large
\textbf{Hyperfine rovibrational states of H$_3^+$ in a weak external magnetic field}
} \\[0.5cm]

Gustavo Avila,$^1$ Ayaki Sunaga,$^1$ Stanislav Komorovsky,$^2$ Edit M\'atyus$^1$ \\
\emph{$^1$~ELTE, E\"otv\"os Lor\'and University, Institute of Chemistry, %
P\'azm\'any P\'eter s\'et\'any 1/A, %
1117 Budapest, Hungary} \\
\emph{$^2$~Institute of Inorganic Chemistry, Slovak Academy of Sciences, Dubravska cesta 9, 84536 Bratislava, Slovakia}
\end{center}

\date{\today}

\setcounter{section}{0}
\renewcommand{\thesection}{S\arabic{section}}
\setcounter{subsection}{0}
\renewcommand{\thesubsection}{S\arabic{section}.\arabic{subsection}}

\setcounter{equation}{0}
\renewcommand{\theequation}{S\arabic{equation}}

\setcounter{table}{0}
\renewcommand{\thetable}{S\arabic{table}}

\setcounter{figure}{0}
\renewcommand{\thefigure}{S\arabic{figure}}


\vspace{0.5cm}
\begin{flushleft}
  S1. Wang functions and symmetric-top eigenfunctions \\
  S2. General second-rank tensors expressed with Wigner $D$ functions \\
  S3. Further computational details and convergence tests \\
  S4. A simple analytic representation of the electric dipole polarizability matrix \\
  S5. Pauli-allowed states: Calculation of the effect of the $P_{23}$ permutation operator
\end{flushleft}

\clearpage

\section{Wang functions and symmetric-top eigenfunctions}
\noindent %
In the GENIUSH program \cite{MaCzCs09,FaMaCs11}, the rovibrational wave function is expressed as 
\begin{align}
  \Psi_n^{(JM)}(\Omega,\bq)
  =
  \sum_{v\wK\tau}
    c_{n,v\wK\tau}^{(JM)}
    f_v(\bq)
    |J\wK\tau M)_\Omega \; ,
\end{align}
where $|J\wK\tau M\rangle$ labels a Wang function,
which is obtained as the following linear combination of the symmetric top functions,
\begin{align}
  |J\wK\tau M \rangle_\Omega
  =
  \left\lbrace 
  \begin{array}{@{}c@{\ }l@{}}
       \frac{1}{\sqrt{2}} [ |J\wK M\rangle_\Omega + |J(-\wK)M\rangle_\Omega ], & \text{for even}~\wK, \tau=0  \\
    \frac{\iim}{\sqrt{2}} [ |J\wK M\rangle_\Omega - |J(-\wK)M\rangle_\Omega ], & \text{for even}~\wK, \tau=1 \\
       \frac{1}{\sqrt{2}} [ |J\wK M\rangle_\Omega - |J(-\wK)M\rangle_\Omega ], & \text{for odd}~\wK,  ~\tau=0\\
    \frac{\iim}{\sqrt{2}} [ |J\wK M\rangle_\Omega + |J(-\wK)M\rangle_\Omega ], & \text{for odd}~\wK,  ~\tau=1 \; \\  
  \end{array}
  \right. \; 
  \label{eq:Wang}
\end{align}
with $M=-J,-J+1,\ldots,J-1,J$, $\wK=0,1,\ldots,J$, $\tau=0,1$. 
For $\wK=0$ there is only $\tau=0$ with $|J,\wK=0,\tau=0, M \rangle_\Omega = |J00 \rangle_\Omega$.

This linear combination of the symmetric top functions ensures that all rovibrational Hamiltonian matrix elements are real. 
The symmetric top eigenfunctions are, $\tK,M=-J,-J+1,\ldots,J-1,J$,
\begin{align}
  |J \tK M\rangle_\Omega 
  =
  \bar{D}^{\text{Z},J\ast}_{M \tK}(\alpha,\beta,\gamma) 
  =
  \sqrt{\frac{2J+1}{8\pi^2}} D^{\text{Z},J\ast}_{M \tK}(\alpha,\beta,\gamma) \; ,
  \label{eq:WignerDZ}
\end{align}
and we also note that
\begin{align}
  \bar{D}^{\text{Z},J}_{M,\tK}(\Omega) 
  =
  (-1)^{M-\tK}\bar{D}^{\text{Z},J*}_{-M,-\tK}(\Omega) \; .
\end{align}

\section{General second-rank tensors expressed with Wigner $D$ functions}
The laboratory-frame property tensors are connected to their body-fixed frame versions 
through transformation with the rotation matrix defined as \cite{Za98}
\begin{align}
  \bos{O}(\alpha,\beta,\gamma)
  =
  \bos{O}_1(\alpha)
  \bos{O}_2(\beta)
  \bos{O}_3(\gamma)
  \label{eq:rotmx1}
\end{align}
with 
{\footnotesize%
\begin{align}
  \bos{O}_1(\alpha) 
  =
  \left(%
  \begin{array}{@{}rrr@{}}
     \cos\alpha & -\sin\alpha & 0 \\
     \sin\alpha &  \cos\alpha & 0 \\
              0 &          0 & 1 
  \end{array}
  \right) 
  ,\ \ 
  \bos{O}_2(\beta)
  =
  \left(%
  \begin{array}{@{}ccc@{}}
    \cos\beta & 0 & \sin\beta \\
            0 & 1 & 0 \\
    -\sin\beta & 0 & \cos\beta 
  \end{array}
  \right)
  ,\ \ 
  \bos{O}_3(\gamma) 
  =
  \left(%
  \begin{array}{@{}rrr@{}}
     \cos\gamma & -\sin\gamma & 0 \\
     \sin\gamma &  \cos\gamma & 0 \\
              0 &          0 & 1 
  \end{array}  
  \right) \; .
  \label{eq:rotmx2}
\end{align}
}
Then,
\begin{align}
  \bos{T}(\Omega,\bq) 
  =
  \bos{O}(\Omega)
  \bos{t}(\bq)
  \bos{O}^\tT(\Omega) \; .
\end{align}
This orientational dependence of a second-rank tensor is conveniently expressed using the Wigner~   $D$ functions as
\begin{align}
  T_{AB}(\Omega,\bq) 
  =
  \sum_{l=0}^2
    \sum_{p=-l}^l
    \sum_{q=-l}^l
      \bar{D}^{\text{Z},l}_{pq}(\Omega) 
      \barT^{AB}_{lpq}(\bq) \; ,
  \label{eq:tensdec}
\end{align}
where $A,B=X,Y,Z$ label laboratory-frame coordinates 
and we use the normalized Wigner $D$ functions, $\bar{D}^{\ZARE,J}_{M,\tK}(\Omega)$.
In Eq.~\eqref{eq:tensdec}, $\bar{T}^{AB}_{lpq}(\bq)$ can be expressed as a linear combination of 
the body-fixed $t_{ij}(\bq)$ values. These expressions are known in the literature, but different phase conventions exist, so for full consistency within our work, we calculated the decomposition terms with Wolfram Mathematica \cite{WolframMath}
by exploiting the orthonormality of the $\bar{D}^{\ZARE,J}_{M,\tK}$ functions,
\begin{align}
  \barT^{AB}_{lpq}(\bq)
  &=
  \int\limits_{0}^{2\pi} \dd\alpha
  \int\limits_{0}^{\pi} \sin\beta\ \dd\beta
  \int\limits_{0}^{2\pi} \dd\gamma\ 
    \bar{D}^{\text{Z},l\ast}_{pq}(\alpha,\beta,\gamma) T_{AB}(\alpha,\beta,\gamma,\bq)
  \nonumber \\
  &=
  \int\limits_{0}^{2\pi} \dd\alpha
  \int\limits_{0}^{\pi} \sin\beta\ \dd\beta
  \int\limits_{0}^{2\pi} \dd\gamma\ 
    \bar{D}^{\text{Z},l\ast}_{pq}(\alpha,\beta,\gamma)   
    \left[
      \bos{O}(\alpha,\beta,\gamma)
      \bos{t}(\bq)
      \bos{O}^\tT(\alpha,\beta,\gamma)
    \right]_{AB}    
  \nonumber \\
  &=
  \int\limits_{0}^{2\pi} \dd\alpha
  \int\limits_{0}^{\pi} \sin\beta\ \dd\beta
  \int\limits_{0}^{2\pi} \dd\gamma\ 
    \bar{D}^{\text{Z},l\ast}_{pq}(\alpha,\beta,\gamma) 
    \sum_{a,b}
      O_{Aa}(\alpha,\beta,\gamma)
      t_{ab}(\bq)
      O_{Bb}(\alpha,\beta,\gamma) \; .
\end{align}
We also note that the phase convention in Zare's book (indicated with the `Z' superscript) \cite{Za98} is related to the Wigner $D$ function implementation 
in Wolfram Mathematica (indicated with the `M' superscript) as
\begin{align}
  D^{\ZARE,J}_{M,\tK}(\alpha,\beta,\gamma)
  =
  (-1)^{M-\tK} D^{\text{M},J}_{\tK,M}(-\alpha,\beta,-\gamma) \; .
\end{align}
For the sake of completeness, the result of this calculation for a general (non-symmetric) $T$ tensor is listed as follows. We used these expressions in our Fortran implementation. The expressions can be simplified for a symmetric tensor (\emph{e.g.,} the polarizability tensor).
\begin{align}
\barT^{X,X}_{0,0,0}   &= (2\sqrt{2}(t_{x,x}^{\MF}+t_{y,y}^{\MF}+t_{z,z}^{\MF})\pi)/3 \nonumber \\
\barT^{X,X}_{2,-2,-2} &= ((t_{x,x}^{\MF} + \iim ( t_{x,y}^{\MF}+t_{y,x}^{\MF}+t_{x,x}^{\MF}+\iim  t_{y,y}^{\MF}))\pi)/\sqrt{10} \nonumber\\
\barT^{X,X}_{2,-2,-1} &= ((t_{x,z}^{\MF} +\iim  t_{y,z}^{\MF} + t_{z,x}^{\MF} +\iim t_{z,y}^{\MF})\pi)/\sqrt{10} \nonumber\\
\barT^{X,X}_{2,-2,0}  &= -(((t_{x,x}^{\MF} + t_{y,y}^{\MF} - 2t_{z,z}^{\MF})\pi)/\sqrt{15}) \nonumber\\
\barT^{X,X}_{2,-2,1 } &= -(((t_{x,z}^{\MF} -\iim (t_{y,z}^{\MF} +\iim t_{z,x}^{\MF} + t_{z,y}^{\MF}))\pi)/\sqrt{10}) \nonumber\\
\barT^{X,X}_{2,-2,2}  &= ((t_{x,x}^{\MF} -\iim (t_{x,y}^{\MF} + t_{y,x}^{\MF} -\iim t_{y,y}^{\MF}))\pi)/\sqrt{10}\nonumber\\
\barT^{X,X}_{2,0,-2}  &= -(((t_{x,x}^{\MF} +\iim (t_{x,y}^{\MF} + t_{y,x}^{\MF} +\iim t_{y,y}^{\MF}))\pi)/\sqrt{15}) \nonumber\\
\barT^{X,X}_{2,0,-1}  &= -(((t_{x,z}^{\MF} +\iim t_{y,z}^{\MF} + t_{z,x}^{\MF} +\iim t_{z,y}^{\MF})\pi)/\sqrt{15}) \nonumber\\
\barT^{X,X}_{2,0, 0}  &= (\sqrt{2/5}(t_{x,x}^{\MF} + t_{y,y}^{\MF} - 2t_{z,z}^{\MF})\pi)/3 \nonumber\\
\barT^{X,X}_{2,0, 1}  &= ((t_{x,z}^{\MF} -\iim (t_{y,z}^{\MF} +\iim t_{z,x}^{\MF} + t_{z,y}^{\MF})))/\sqrt{15} \nonumber\\
\barT^{X,X}_{2,0, 2}  &= ((-t_{x,x}^{\MF} +\iim (t_{x,y}^{\MF} + t_{y,x}^{\MF}) + t_{y,y}^{\MF})\pi)/\sqrt{15} \nonumber \\
\barT^{X,X}_{2,2,-2}  &= ((t_{x,x}^{\MF} +\iim (t_{x,y}^{\MF} + t_{y,x}^{\MF} +\iim t_{y,y}^{\MF}))\pi)/\sqrt{10} \nonumber\\
\barT^{X,X}_{2,2,-1}  &= ((t_{x,z}^{\MF} +\iim t_{y,z}^{\MF} + t_{z,x}^{\MF} +\iim t_{z,y}^{\MF})\pi)/\sqrt{10} \nonumber\\
\barT^{X,X}_{2,2, 0}  &= -(((t_{x,x}^{\MF} + t_{y,y}^{\MF} - 2t_{z,z})\pi)/\sqrt{15}) \nonumber\\
\barT^{X,X}_{2,2, 1}  &= -(((t_{x,z}^{\MF} -\iim (t_{y,z}^{\MF} +\iim t_{z,x}^{\MF} + t_{z,y}^{\MF}))\pi)/\sqrt{10}) \nonumber\\
\barT^{X,X}_{2,-2, 2} &= ((t_{x,x}^{\MF} -\iim (t_{x,y}^{\MF} + t_{y,x}^{\MF}) - t_{y,y}^{\MF})\pi)/\sqrt{10} 
\end{align}
\begin{align}
\barT^{X,Y}_{1,0,-1}  &= (( -\iim t_{x,z}^{\MF} + t_{y,z}^{\MF} + \iim t_{z,x}^{\MF} - t_{z,y}^{\MF})\pi)/\sqrt{3}         \nonumber  \\
\barT^{X,Y}_{1,0,0}   &= \sqrt{2/3}(t_{x,y}^{\MF} - t_{y,x}^{\MF})\pi                                              \nonumber  \\
\barT^{X,Y}_{1,0,1}   &= (( -\iim t_{x,z}^{\MF} - t_{y,z}^{\MF} + \iim t_{z,x}^{\MF} + t_{z,y}^{\MF}) \pi )/\sqrt{3}          \nonumber  \\
\barT^{X,Y}_{2,-2,-2} &=  (( -\iim t_{x,x}^{\MF} + t_{x,y}^{\MF} + t_{y,x}^{\MF} + \iim t_{y,y}^{\MF}) \pi )/\sqrt{10}      \nonumber  \\
\barT^{X,Y}_{2,-2,-1} &=  (( -\iim t_{x,z}^{\MF} + t_{y,z}^{\MF} - \iim t_{z,x}^{\MF} + t_{z,y}^{\MF}) \pi )/\sqrt{10}      \nonumber  \\
\barT^{X,Y}_{2,-2,0}  &=  (\iim (t_{x,x}^{\MF} + t_{y,y}^{\MF} - 2t_{z,z}^{\MF}) \pi )/\sqrt{15}                         \nonumber  \\
\barT^{X,Y}_{2,-2,1}  &=  ((\iim t_{x,z}^{\MF} + t_{y,z}^{\MF} + \iim t_{z,x}^{\MF} + t_{z,y}^{\MF}) \pi )/\sqrt{10}           \nonumber  \\
\barT^{X,Y}_{2,-2,2}  &=  ( -\iim (t_{x,x}^{\MF} - \iim (t_{x,y}^{\MF} + t_{y,x}^{\MF}) - t_{y,y}^{\MF}) \pi )/\sqrt{10}     \nonumber  \\
\barT^{X,Y}_{2,2,-2}  &=  -((( -\iim t_{x,x}^{\MF} + t_{x,y}^{\MF} + t_{y,x}^{\MF} + \iim t_{y,y}^{\MF}) \pi )/\sqrt{10})    \nonumber  \\
\barT^{X,Y}_{2,2,-1}  &=  (i(t_{x,z}^{\MF} + \iim t_{y,z}^{\MF} + t_{z,x}^{\MF} + \iim t_{z,y}^{\MF}) \pi )/\sqrt{10}          \nonumber  \\
\barT^{X,Y}_{2,2,0}   &=  ( -\iim (t_{x,x}^{\MF} + t_{y,y}^{\MF} - 2t_{z,z}^{\MF}) \pi )/\sqrt{15}                      \nonumber  \\
\barT^{X,Y}_{2,2,1}   &=  ( -\iim t_{x,z}^{\MF}\pi - (t_{y,z}^{\MF} + \iim t_{z,x}^{\MF} + t_{z,y}^{\MF}) \pi )/\sqrt{10}     \nonumber  \\
\barT^{X,Y}_{2,2,2}   &=  ((\iim t_{x,x}^{\MF} + t_{x,y}^{\MF} + t_{y,x}^{\MF} - \iim t_{y,y}^{\MF}) \pi )/\sqrt{10}                 
\end{align}
\begin{align}
\barT^{X,Z}_{1,-1,-1} &= ((t_{x,z}^{\MF} + \iim (t_{y,x}^{\MF} + \iim t_{z,x}^{\MF} - t_{z,y}^{\MF}))\pi)/\sqrt{6}            \nonumber \\
\barT^{X,Z}_{1,-1,0}  &= ( \iim (t_{x,y}^{\MF} - t_{y,x}^{\MF})\pi)/\sqrt{3}                                              \nonumber \\
\barT^{X,Z}_{1,-1,1}  &= ((t_{x,z}^{\MF} - \iim t_{y,x}^{\MF} - t_{z,x}^{\MF} + \iim t_{z,y}^{\MF})\pi)/\sqrt{6}               \nonumber \\
\barT^{X,Z}_{1,1,-1}  &=   ((t_{x,z}^{\MF} + \iim (t_{y,z}^{\MF} + \iim t_{z,x}^{\MF} - t_{z,y}^{\MF}))\pi)/\sqrt{6}          \nonumber \\
\barT^{X,Z}_{1,1, 0}  &=   ( \iim (t_{x,y}^{\MF} - t_{y,x}^{\MF})\pi)/\sqrt{3}                                           \nonumber \\
\barT^{X,Z}_{1,1, 1}  &=    ((t_{x,z}^{\MF} - \iim t_{y,z}^{\MF} - t_{z,x}^{\MF} + \iim t_{z,y}^{\MF})\pi)/\sqrt{6}           \nonumber \\
\barT^{X,Z}_{2,-1,-2} &=    ((t_{x,x}^{\MF} + \iim (t_{x,y}^{\MF} + t_{y,x}^{\MF} + \iim t_{y,y}^{\MF}))\pi)/\sqrt{10}       \nonumber \\
\barT^{X,Z}_{2,-1,-1} &=    ((t_{x,z}^{\MF} + \iim t_{y,z}^{\MF} + t_{z,x}^{\MF} + \iim t_{z,y}^{\MF})\pi)/\sqrt{10}         \nonumber \\
\barT^{X,Z}_{2,-1, 0} &=    -(((t_{x,x}^{\MF} + t_{y,y}^{\MF} - 2 t_{z,z}^{\MF})\pi)/\sqrt{15})                     \nonumber \\
\barT^{X,Z}_{2,-1, 1} &=    -(((t_{x,z}^{\MF} - \iim (t_{y,z}^{\MF} + \iim t_{z,x}^{\MF} + t_{z,y}^{\MF}))\pi)/\sqrt{10})    \nonumber \\
\barT^{X,Z}_{2,-1, 2} &=    ((t_{x,x}^{\MF} - \iim (t_{x,y}^{\MF} + t_{y,x}^{\MF} - \iim t_{y,y}^{\MF}))\pi)/\sqrt{10}       \nonumber \\
\barT^{X,Z}_{1,1,-2}  &=    -(((t_{x,x}^{\MF} + \iim (t_{x,y}^{\MF} + t_{y,x}^{\MF} + \iim t_{y,y}^{\MF}))\pi)/\sqrt{10})     \nonumber \\
\barT^{X,Z}_{1,1,-1}  &=    -(((t_{x,z}^{\MF} + \iim t_{y,z}^{\MF} + t_{z,x}^{\MF} + \iim t_{z,y}^{\MF})\pi)/\sqrt{10})       \nonumber \\
\barT^{X,Z}_{1,1, 0}  &=    ((t_{x,x}^{\MF} + t_{y,y}^{\MF} - 2 t_{z,z}^{\MF})\pi)/\sqrt{15}                         \nonumber \\
\barT^{X,Z}_{1,1, 1}  &=    ((t_{x,z}^{\MF} - \iim (t_{y,z}^{\MF} + \iim t_{z,x}^{\MF} + t_{z,y}^{\MF}))\pi)/\sqrt{10}        \nonumber \\
\barT^{X,Z}_{1,1, 2}  &=    ((-t_{x,x}^{\MF} + \iim (t_{x,y}^{\MF} + t_{y,x}^{\MF}) + t_{y,y}^{\MF})\pi)/\sqrt{10}       
\end{align}
\begin{align}
\barT^{Y,X}_{1,0,-1} &=  (( \iim t_{x,z}^{\MF} - t_{y,z}^{\MF} - \iim t_{z,x}^{\MF} + t_{z,y}^{\MF})\pi)/\sqrt{3}           \nonumber \\
\barT^{Y,X}_{1,0, 0} &=  \sqrt{2/3} (-t_{x,y}^{\MF} + t_{y,x}^{\MF})\pi                                          \nonumber \\
\barT^{Y,X}_{1,0, 1} &=  (( \iim t_{x,z}^{\MF} + t_{y,z}^{\MF} - \iim t_{z,x}^{\MF} - t_{z,y}^{\MF})\pi)/\sqrt{3}           \nonumber \\
\barT^{Y,X}_{2,-2,-2} &=  (( -\iim  t_{x,x}^{\MF} + t_{x,y}^{\MF} + t_{y,x}^{\MF} + \iim t_{y,y}^{\MF})\pi)/\sqrt{10}     \nonumber \\
\barT^{Y,X}_{2,-2,-1} &=  (( -\iim  t_{x,z}^{\MF} + t_{y,z}^{\MF} - \iim t_{z,x}^{\MF} + t_{z,y}^{\MF})\pi)/\sqrt{10}     \nonumber \\
\barT^{Y,X}_{2,-2,0} &=  ( \iim (t_{x,x}^{\MF} + t_{y,y}^{\MF} - 2 t_{z,z}^{\MF})\pi)/\sqrt{15}                        \nonumber \\
\barT^{Y,X}_{2,-2,1} &=  (( \iim t_{x,z}^{\MF} + t_{y,z}^{\MF} + \iim t_{z,x}^{\MF} + t_{z,y}^{\MF})\pi)/\sqrt{10}          \nonumber \\
\barT^{Y,X}_{2,-2,2} &=  ( -\iim  (t_{x,x}^{\MF} - i(t_{x,y}^{\MF} + t_{y,x}^{\MF}) - t_{y,y}^{\MF})\pi)/\sqrt{10}    \nonumber \\
\barT^{Y,X}_{2,2,-1} &=  -((( -\iim  t_{x,x}^{\MF} + t_{x,y}^{\MF} + t_{y,x}^{\MF} + \iim t_{y,y}^{\MF})\pi)/\sqrt{10})   \nonumber \\
\barT^{Y,X}_{2,2,-1} &=  ( \iim (t_{x,z}^{\MF} + \iim t_{y,z}^{\MF} + t_{z,x}^{\MF} + \iim t_{z,y}^{\MF})\pi)/\sqrt{10}         \nonumber \\
\barT^{Y,X}_{2,2,-1} &=  ( -\iim  (t_{x,x}^{\MF} + t_{y,y}^{\MF} - 2 t_{z,z}^{\MF})\pi)/\sqrt{15}                    \nonumber \\
\barT^{Y,X}_{2,2,-1} &=  ( -\iim  t_{x,z}^{\MF}\pi - (t_{y,z}^{\MF} + \iim t_{z,x}^{\MF} + t_{z,y}^{\MF})\pi)/\sqrt{10}   \nonumber \\
\barT^{Y,X}_{2,2,-1} &=  (( \iim t_{x,x}^{\MF} + t_{x,y}^{\MF} + t_{y,x}^{\MF} - \iim t_{y,y}^{\MF})\pi)/\sqrt{10}         
\end{align}
\begin{align}
\barT^{Y,Y}_{0,0,0}    &=  (2\sqrt{2} (t_{x,x}^{\MF}  + t_{y,y}^{\MF}  + t_{z,z}^{\MF} )\pi)/3  \nonumber \\
\barT^{Y,Y}_{2,-2,-2}  &=      -(((t_{x,x}^{\MF} + \iim (t_{x,y}^{\MF} + t_{y,x}^{\MF} + \iim t_{y,y}^{\MF}))\pi)/\sqrt{10}) \nonumber \\
\barT^{Y,Y}_{2,-2,-1}  &=      -(((t_{x,z}^{\MF} + \iim t_{y,z}^{\MF} + t_{z,x}^{\MF} + \iim t_{z,y}^{\MF})\pi)/\sqrt{10})   \nonumber \\
\barT^{Y,Y}_{2,-2, 0}  &=      ((t_{x,x}^{\MF} + t_{y,y}^{\MF} - 2 t_{z,z}^{\MF})\pi)/\sqrt{15}                     \nonumber \\
\barT^{Y,Y}_{2,-2, 1}  &=      ((t_{x,z}^{\MF} - \iim (t_{y,z}^{\MF} + \iim t_{z,x}^{\MF} + t_{z,y}^{\MF}))\pi)/\sqrt{10}    \nonumber \\
\barT^{Y,Y}_{2,-2, 2}  &=      -(((t_{x,x}^{\MF} - \iim (t_{x,y}^{\MF} + t_{y,x}^{\MF}) - t_{y,y}^{\MF})\pi)/\sqrt{10})  \nonumber \\
\barT^{Y,Y}_{2,0,-2}   &=      -(((t_{x,x}^{\MF} + \iim (t_{x,y}^{\MF} + t_{y,x}^{\MF} + \iim t_{y,y}^{\MF}))\pi)/\sqrt{15})  \nonumber \\
\barT^{Y,Y}_{2,0,-1}   &=      -(((t_{x,z}^{\MF} + \iim t_{y,z}^{\MF} + t_{z,x}^{\MF} + \iim t_{z,y}^{\MF})\pi)/\sqrt{15})    \nonumber \\
\barT^{Y,Y}_{2,0, 0}   &=      (\sqrt{2/5} (t_{x,x}^{\MF} + t_{y,y}^{\MF} - 2 t_{z,z}^{\MF})\pi)/3                   \nonumber \\
\barT^{Y,Y}_{2,0, 1}   &=      ((t_{x,z}^{\MF} - \iim (t_{y,z}^{\MF} + \iim t_{z,x}^{\MF} + t_{z,y}^{\MF}))\pi)/\sqrt{15}     \nonumber \\
\barT^{Y,Y}_{2,0, 2}   &=      -(((t_{x,x}^{\MF} - \iim (t_{x,y}^{\MF} + t_{y,x}^{\MF}) - t_{y,y}^{\MF})\pi)/\sqrt{15})   \nonumber \\
\barT^{Y,Y}_{2,2,-1}   &=      -(((t_{x,x}^{\MF} + \iim (t_{x,y}^{\MF} + t_{y,x}^{\MF} + \iim t_{y,y}^{\MF}))\pi)/\sqrt{10})  \nonumber \\
\barT^{Y,Y}_{2,2,-2}   &=      -(((t_{x,z}^{\MF} + \iim t_{y,z}^{\MF} + t_{z,x}^{\MF} + \iim t_{z,y}^{\MF})\pi)/\sqrt{10})    \nonumber \\
\barT^{Y,Y}_{2,2, 0}   &=      ((t_{x,x}^{\MF} + t_{y,y}^{\MF} - 2 t_{z,z}^{\MF})\pi)/\sqrt{15}                      \nonumber \\
\barT^{Y,Y}_{2,2, 1}   &=      ((t_{x,z}^{\MF} - \iim (t_{y,z}^{\MF} + \iim t_{z,x}^{\MF} + t_{z,y}^{\MF}))\pi)/\sqrt{10}     \nonumber \\
\barT^{Y,Y}_{2,2, 2}   &=      ((-t_{x,x}^{\MF} + \iim (t_{x,y}^{\MF} + t_{y,x}^{\MF}) + t_{y,y}^{\MF})\pi)/\sqrt{10}     
\end{align}
\begin{align}
\barT^{Y,Z}_{2,-1,-1}   &=   (( -\iim  t_{x,z}^{\MF} + t_{y,z}^{\MF} + \iim t_{z,x} - t_{z,y}) \pi)/\sqrt{6}                         \nonumber \\
\barT^{Y,Z}_{2,-1, 0}   &=   ((t_{x,y}^{\MF} - t_{y,x}^{\MF}) \pi)/\sqrt{3}                                           \nonumber \\
\barT^{Y,Z}_{2,-1, 1}   &=   (( -\iim  t_{x,z}^{\MF} - t_{y,z}^{\MF} + \iim t_{z,x} + t_{z,y}) \pi)/\sqrt{6}                         \nonumber \\
\barT^{Y,Z}_{1,1,-1}   &=   ((\iim t_{x,z}^{\MF} - t_{y,z}^{\MF} - \iim t_{z,x} + t_{z,y}) \pi)/\sqrt{6}                              \nonumber \\
\barT^{Y,Z}_{1,1, 0}   &=   ((-t_{x,y}^{\MF} + t_{y,x}^{\MF}) \pi)/\sqrt{3}                                           \nonumber \\
\barT^{Y,Z}_{1,1, 1}   &=    ((\iim t_{x,z}^{\MF} + t_{y,z}^{\MF} - \iim t_{z,x} - t_{z,y}) \pi)/\sqrt{6}                             \nonumber \\
\barT^{Y,Z}_{2,-1,-2}   &=    (( -\iim  t_{x,x}^{\MF} + t_{x,y}^{\MF} + t_{y,x}^{\MF} + \iim t_{y,y}^{\MF}) \pi)/\sqrt{10}     \nonumber \\
\barT^{Y,Z}_{2,-1,-1}   &=    (( -\iim  t_{x,z}^{\MF} + t_{y,z}^{\MF} - \iim t_{z,x} + t_{z,y}) \pi)/\sqrt{10}                       \nonumber \\
\barT^{Y,Z}_{2,-1, 0}   &=    ( \iim (t_{x,x}^{\MF} + t_{y,y}^{\MF} - 2 t_{z,z}^{\MF}) \pi)/\sqrt{15}                       \nonumber \\
\barT^{Y,Z}_{2,-1, 1}   &=    ((\iim t_{x,z}^{\MF} + t_{y,z}^{\MF} + \iim t_{z,x} + t_{z,y}) \pi)/\sqrt{10}                           \nonumber \\
\barT^{Y,Z}_{2,-1, 2}   &=    ( -\iim  (t_{x,x}^{\MF} -  \iim (t_{x,y}^{\MF} + t_{y,x}^{\MF} - \iim t_{y,y}^{\MF})) \pi)/\sqrt{10}  \nonumber \\
\barT^{Y,Z}_{2,1,-2}   &=    (( -\iim  t_{x,x}^{\MF} + t_{x,y}^{\MF} + t_{y,x}^{\MF} + \iim t_{y,y}^{\MF}) \pi)/\sqrt{10}      \nonumber \\
\barT^{Y,Z}_{2,1,-1}   &=    (( -\iim  t_{x,z}^{\MF} + t_{y,z}^{\MF} - \iim t_{z,x} + t_{z,y}) \pi)/\sqrt{10}                        \nonumber \\
\barT^{Y,Z}_{2,1, 0}   &=    ( \iim (t_{x,x}^{\MF} + t_{y,y}^{\MF} - 2 t_{z,z}^{\MF}) \pi)/\sqrt{15}                        \nonumber \\
\barT^{Y,Z}_{2,1, 1}   &=    ((\iim t_{x,z}^{\MF} + t_{y,z}^{\MF} + \iim t_{z,x} + t_{z,y}) \pi)/\sqrt{10}                            \nonumber \\
\barT^{Y,Z}_{2,1, 2}   &=    ( -\iim  (t_{x,x}^{\MF} -  \iim (t_{x,y}^{\MF} + t_{y,x}^{\MF}) - t_{y,y}^{\MF}) \pi)/\sqrt{10}    
\end{align}
\begin{align}
\barT^{Z,X}_{1,-1,-1}   &=    ((-t_{x,z}^{\MF} -\iim t_{y,z}^{\MF} + t_{z,x}^{\MF} +\iim t_{z,y}^{\MF})\pi)/\sqrt{6}         \nonumber \\
\barT^{Z,X}_{1,-1, 0}   &=    ( -\iim  (t_{x,y}^{\MF} - t_{y,x}^{\MF})\pi)/\sqrt{3}                                     \nonumber \\
\barT^{Z,X}_{1,-1, 1}   &=    ((-t_{x,z}^{\MF} +\iim t_{y,z}^{\MF} + t_{z,x}^{\MF} -\iim t_{z,y}^{\MF})\pi)/\sqrt{6}         \nonumber \\
\barT^{Z,X}_{1, 1,-1}   &=    ((-t_{x,z}^{\MF} -\iim t_{y,z}^{\MF} + t_{z,x}^{\MF} +\iim t_{z,y}^{\MF})\pi)/\sqrt{6}         \nonumber \\
\barT^{Z,X}_{1, 1, 0}   &=    ( -\iim  (t_{x,y}^{\MF} - t_{y,x}^{\MF})\pi)/\sqrt{3}                                     \nonumber \\
\barT^{Z,X}_{1, 1, 1}   &=     ((-t_{x,z}^{\MF} +\iim t_{y,z}^{\MF} + t_{z,x}^{\MF} -\iim t_{z,y}^{\MF})\pi)/\sqrt{6}        \nonumber \\
\barT^{Z,X}_{2,-1,-2}   &=     ((t_{x,x}^{\MF} +  \iim (t_{x,y}^{\MF} + t_{y,x}^{\MF} +\iim t_{y,y}^{\MF}))\pi)/\sqrt{10}      \nonumber \\
\barT^{Z,X}_{2,-1,-1}   &=     ((t_{x,z}^{\MF} +\iim t_{y,z}^{\MF} + t_{z,x}^{\MF} +\iim t_{z,y}^{\MF})\pi)/\sqrt{10}        \nonumber \\
\barT^{Z,X}_{2,-1, 0}   &=     -(((t_{x,x}^{\MF} + t_{y,y}^{\MF} - 2 t_{z,z}^{\MF})\pi)/\sqrt{15})                    \nonumber \\
\barT^{Z,X}_{2,-1, 1}   &=     -(((t_{x,z}^{\MF} -  \iim (t_{y,z}^{\MF} +\iim t_{z,x}^{\MF} + t_{z,y}^{\MF}))\pi)/\sqrt{10})   \nonumber \\
\barT^{Z,X}_{2,-1, 2}   &=     ((t_{x,x}^{\MF} -  \iim (t_{x,y}^{\MF} + t_{y,x}^{\MF} -\iim t_{y,y}^{\MF}))\pi)/\sqrt{10}      \nonumber \\
\barT^{Z,X}_{2, 1,-2}   &=     -(((t_{x,x}^{\MF} +  \iim (t_{x,y}^{\MF} + t_{y,x}^{\MF} +\iim t_{y,y}^{\MF}))\pi)/\sqrt{10})   \nonumber \\
\barT^{Z,X}_{2, 1,-1}   &=     -(((t_{x,z}^{\MF} +\iim t_{y,z}^{\MF} + t_{z,x}^{\MF} +\iim t_{z,y}^{\MF})\pi)/\sqrt{10})     \nonumber \\
\barT^{Z,X}_{2, 1, 0}   &=     ((t_{x,x}^{\MF} + t_{y,y}^{\MF} - 2 t_{z,z}^{\MF})\pi)/\sqrt{15}                       \nonumber \\
\barT^{Z,X}_{2, 1, 1}   &=     ((t_{x,z}^{\MF} -  \iim (t_{y,z}^{\MF} +\iim t_{z,x}^{\MF} + t_{z,y}^{\MF}))\pi)/\sqrt{10}      \nonumber \\
\barT^{Z,X}_{2, 1, 2}   &=     ((-t_{x,x}^{\MF} +  \iim (t_{x,y}^{\MF} + t_{y,x}^{\MF}) + t_{y,y}^{\MF})\pi)/\sqrt{10}      
\end{align} 
\begin{align}
\barT^{Z,Y}_{1,-1,-1} &=  ((\iim t_{x,z}^{\MF} - t_{y,z}^{\MF} - \iim t_{z,x}^{\MF} + t_{z,y}^{\MF})\pi)/\sqrt{6}          \nonumber \\
\barT^{Z,Y}_{1,-1, 0} &=  ((-t_{x,y}^{\MF} + t_{y,x}^{\MF})\pi)/\sqrt{3}                                         \nonumber \\
\barT^{Z,Y}_{1,-1, 1} &=  ((\iim t_{x,z}^{\MF} + t_{y,z}^{\MF} - \iim t_{z,x}^{\MF} - t_{z,y}^{\MF})\pi)/\sqrt{6}          \nonumber \\
\barT^{Z,Y}_{1, 1,-1} &=  (( -\iim  t_{x,z}^{\MF} + t_{y,z}^{\MF} + \iim t_{z,x}^{\MF} - t_{z,y}^{\MF})\pi)/\sqrt{6}      \nonumber \\
\barT^{Z,Y}_{1, 1, 0} &=  ((t_{x,y}^{\MF} - t_{y,x}^{\MF})\pi)/\sqrt{3}                                          \nonumber \\
\barT^{Z,Y}_{1, 1, 1} &=   (( -\iim  t_{x,z}^{\MF} - t_{y,z}^{\MF} + \iim t_{z,x}^{\MF} + t_{z,y}^{\MF})\pi)/\sqrt{6}     \nonumber \\
\barT^{Z,Y}_{2,-1,-2} &=   (( -\iim  t_{x,x}^{\MF} + t_{x,y}^{\MF} + t_{y,x}^{\MF} + \iim t_{y,y}^{\MF})\pi)/\sqrt{10}    \nonumber \\
\barT^{Z,Y}_{2,-1,-1} &=   (( -\iim  t_{x,z}^{\MF} + t_{y,z}^{\MF} - \iim t_{z,x}^{\MF} + t_{z,y}^{\MF})\pi)/\sqrt{10}    \nonumber \\
\barT^{Z,Y}_{2,-1, 0} &=   ( \iim (t_{x,x}^{\MF} + t_{y,y}^{\MF} - 2 t_{z,z}^{\MF})\pi)/\sqrt{15}                      \nonumber \\
\barT^{Z,Y}_{2,-1, 1} &=   ((\iim t_{x,z}^{\MF} + t_{y,z}^{\MF} + \iim t_{z,x}^{\MF} + t_{z,y}^{\MF})\pi)/\sqrt{10}        \nonumber \\
\barT^{Z,Y}_{2,-1, 2} &=   ( -\iim  (t_{x,x}^{\MF} -  \iim (t_{x,y}^{\MF} + t_{y,x}^{\MF} - \iim t_{y,y}^{\MF}))\pi)/\sqrt{10} \nonumber \\
\barT^{Z,Y}_{2, 1,-2} &=   (( -\iim  t_{x,x}^{\MF} + t_{x,y}^{\MF} + t_{y,x}^{\MF} + \iim t_{y,y}^{\MF})\pi)/\sqrt{10}    \nonumber \\
\barT^{Z,Y}_{2, 1,-1} &=   (( -\iim  t_{x,z}^{\MF} + t_{y,z}^{\MF} - \iim t_{z,x}^{\MF} + t_{z,y}^{\MF})\pi)/\sqrt{10}    \nonumber \\
\barT^{Z,Y}_{2, 1, 0} &=   ( \iim (t_{x,x}^{\MF} + t_{y,y}^{\MF} - 2 t_{z,z}^{\MF})\pi)/\sqrt{15}                      \nonumber \\
\barT^{Z,Y}_{2, 1, 1} &=   ((\iim t_{x,z}^{\MF} + t_{y,z}^{\MF} + \iim t_{z,x}^{\MF} + t_{z,y}^{\MF})\pi)/\sqrt{10}        \nonumber \\
\barT^{Z,Y}_{2, 1, 2} &=   ( -\iim  (t_{x,x}^{\MF} -  \iim (t_{x,y}^{\MF} + t_{y,x}^{\MF}) - t_{y,y}^{\MF})\pi)/\sqrt{10}  
\end{align}
\begin{align}
\barT^{Z,Z}_{0, 0, 0} &=    (2 \sqrt{2} (t_{x,x}^{\MF} + t_{y,y}^{\MF} + t_{z,z}^{\MF})\pi)/3                         \nonumber \\
\barT^{Z,Z}_{2,0,-2}  &=     (2 (t_{x,x}^{\MF} + \iim (t_{x,y}^{\MF} + t_{y,x}^{\MF} + \iim t_{y,y}^{\MF}))\pi)/\sqrt{15}       \nonumber \\
\barT^{Z,Z}_{2,0,-1}  &=     (2 (t_{x,z}^{\MF} + \iim t_{y,z}^{\MF} + t_{z,x}^{\MF} + \iim t_{z,y}^{\MF})\pi)/\sqrt{15}         \nonumber \\
\barT^{Z,Z}_{2,0, 0}  &=     (-2 \sqrt{2/5} (t_{x,x}^{\MF} + t_{y,y}^{\MF} - 2 t_{z,z}^{\MF})\pi)/3                    \nonumber \\
\barT^{Z,Z}_{2,0, 1}  &=     (-2 (t_{x,z}^{\MF} - \iim (t_{y,z}^{\MF} + \iim t_{z,x}^{\MF} + t_{z,y}^{\MF}))\pi)/\sqrt{15}      \nonumber \\
\barT^{Z,Z}_{2,0, 2}  &=     (2 (t_{x,x}^{\MF} - \iim (t_{x,y}^{\MF} + t_{y,x}^{\MF} - \iim t_{y,y}^{\MF}))\pi)/\sqrt{15}     
\end{align}

\vspace{0.5cm}
During the calculations, the product of three Wigner $D$ functions are integrated according to the known relation, Eq.~3.118 of Ref.~\cite{Za98},
\begin{align}
  \int\limits_{0}^{2\pi} \dd\alpha
  \int\limits_{0}^{\pi} \sin\beta\ \dd\beta
  \int\limits_{0}^{2\pi} \dd\gamma\ 
    D^{\text{Z},J_3}_{M_{3},K_{3}}(\alpha,\beta,\gamma) 
    D^{\text{Z},J_2}_{M_{2},K_{2}}(\alpha,\beta,\gamma) 
    D^{\text{Z},J_1}_{M_{1},K_{1}}(\alpha,\beta,\gamma)
    \nonumber \\
  =
  8\pi^2
  \begin{pmatrix}
     J_{1} & J_{2} & J_{3}  \\
     M_{1} & M_{2} & M_{3}   
  \end{pmatrix}
  \begin{pmatrix}
     J_{1} & J_{2} & J_{3}  \\
     K_{1} & K_{2} & K_{3}   
  \end{pmatrix}~~~~~~~~~~~~~~~~~~~~~
\end{align}
expressed with 3j symbols. 

Furthermore, the action of the laboratory-fixed angular momentum operators on the rigid rotor functions can be calculated (by ladder operators) as
\begin{align}
  \hat{J}_{X} |J\tK M \rangle
  &= 
  \frac{1}{2}\sqrt{J(J+1)-M(M+1)} |J \tK M+1\rangle \nonumber \\
  &\quad %
  +\frac{1}{2}\sqrt{J(J+1)-M(M-1)} |J \tK M-1\rangle 
  \label{eq:JX}
  \\
  \hat{J}_{Y} |J \tK M \rangle
  &= 
  -\frac{\iim}{2}\sqrt{J(J+1)-M(M+1)} |J \tK M+1\rangle \nonumber \\
  &\quad %
  +\frac{\iim}{2}\sqrt{J(J+1)-M(M-1)} |J \tK M-1\rangle
  \label{eq:JY}
  \\
  \hat{J}_{Z} |J \tK M \rangle
  &= 
  M |J \tK M\rangle \; .
  \label{eq:JZ}
\end{align}

\clearpage
\section{Further computational details and convergence tests}
Throughout the computations the proton mass was $m_\text{p}=1.007\ 276\ 466\ 578\ 9$~u, and the proton $g$ factor was $g_\text{p}=5.585\ 694\ 689\ 3$ \cite{codata22}.

\linespread{1.}

\begin{table}[H]
\caption{%
  Rotational energies up to $J=3$ corresponding to the vibrational ground state computed with GENIUSH using the $\rone,\rtwo,\tangle$ valence coordinates and the GLH3P PES \cite{pes12}. 
  For the $\rone$ and $\rtwo$ stretching degrees of freedom, 
  we used PO-DVR constructed over a Laguerre DVR based on the $L^{(\alpha=2)}_n$ associated Laguerre polynomials, and 61 grid points scaled to the $[0.3,6.5]$~bohr interval. 
  For the $\cos\tangle$ angle, we used Jacobi DVR based on the $J^{(\alpha=0.05,\beta=0.05)}_n$ Jacobi polynomials and 91 DVR points.
  This (61,61,91) grid was sufficient to converge the energies to $10^{-6}$~\cm\ (with the present PES).
  The MARVEL compilation of experimental data and the conventional $(J,m)$ labelling of the rovibrational states are also reproduced from Ref.~\cite{FuSzMaFaCs13}.   
  \label{tab:convtest}
}
\centering
\begin{tabular}{@{}cccrr@{}}
 \hline\hline\\[-0.35cm]
   &   &           &  \multicolumn{2}{c}{$\tilde\nu$ [cm$^{-1}$]} \\
   \cline{4-5}\\[-0.35cm]
 $J$ & $m$ & $\Gamma$ & \multicolumn{1}{c}{This work} & \multicolumn{1}{c}{MARVEL \cite{FuSzMaFaCs13}} \\
\hline\\[-0.35cm]
 1 & 1 & $\Epp$    &     64.128293876  &   64.121000 \\
 1 & 0 & $\Atwop$  &     86.966144514  &   86.960000 \\
 2 & 2 & $\Ep$     &    169.308639849  &  169.294000 \\
 2 & 1 & $\Epp$    &    237.368760616  &  237.357000 \\
 3 & 3 & $\Atwop$  &    315.364469199  &  315.354081 \\
 3 & 2 & $\Ep$     &    428.041883665  &  428.019000 \\
 3 & 1 & $\Epp$    &    494.792192357  &  494.773333 \\
 3 & 0 & $\Atwop$  &    516.907378512  &  516.878695 \\
 \hline\hline
\end{tabular}
\end{table}

\begin{table}[htbp]
\caption{%
Basis set dependence of the magnetic properties at the CCSD level: $\gmol$ (dimensionless), isotropic $\srot$ in kHz, and isotropic $\shield$ in ppm ($10^{-6}$). $r_{12} = r_{13}=$ 1.65~bohr and $\theta_{213}=60^\circ$. The molecule is placed in the $yz$ plane with the $z$ axis parallel to the bisector of the H$_2$-H$_1$-H$_3$ angle. } \label{tab:ccsd_convtest}
\begin{tabular}{@{}lcccc@{}}
  \hline\hline\\[-0.35cm]
  & $\gmol_{xx}$ & $\gmol_{yy/zz}$ & $\srot$  & $\shield$ \\
  \hline\\[-0.35cm]
  cc-pVDZ & 0.9791 & 0.9328 & 73.06  & 20.59 \\
  cc-pVTZ & 0.9761 & 0.9249 & 70.79  & 20.55 \\
  aug-cc-pVTZ & 0.9764 & 0.9265 & 70.76  & 20.55 \\
  cc-pVQZ & 0.9764 & 0.9247 &  70.42 &  20.51\\
\hline\hline
\end{tabular}
\end{table}

\begin{table}[H]
\caption{%
  Summary of the prefactors appearing in our computer implementation of Eqs.~(5) and (6) with the coupling tensors taken as printed by CFOUR (or DALTON), the magnetic induction in Gauss (0.1 mT), and computing the rovibrational-hyperfine-magnetic energies in cm$^{-1}$.
  \label{tab:conversion}
}
\centering
\begin{tabular}{@{}lr@{}}
  \hline\hline\\[-0.3cm]
    & coefficients to cm$^{-1}$ \\
  \hline\\[-0.3cm]
    Spin-rotation coupling matrix [kHz] & %
      $-3.335\ 640\ 951\ 982\cdot 10^{-8}$ \\
    Direct spin-spin coupling [bohr$^{-3}$] & %
      $-2.703\ 907\ 337\ 273\cdot 10^{-5}$ \\
    Indirect spin-spin coupling [Hz] & %
      $3.335\ 640\ 951\ 981\cdot 10^{-11}$ \\
    Rotation-magnetic coupling [dimensionless] & %
      $-2.542\ 623\ 411\ 647\cdot 10^{-8}$  \\
    Direct spin-magnetic coupling [dimensionless] & %
     $1.420\ 231\ 808\ 733\cdot 10^{-7}$ \\
    Spin-magnetic coupling [dimensionless] & %
     $-1.420\ 231\ 808\ 733\cdot 10^{-7}$ \\
  \hline\hline
\end{tabular}
\end{table}


\begin{table}[H]
\caption{%
  Character table and irrep labels of the $S_3$ permutational group. The labelling follows that of 
  $D_{3\text{h}}(\text{M})$ \cite{BuJe98}, but since space inversion is not considered, we dropped the ' and '' superscripts from the $D_{3\text{h}}(\text{M})$ irrep labels.
  \label{tab:s3char}
}
\centering
\begin{tabular}{@{}cccc@{}}
  \hline\hline\\[-0.35cm]
  Classes: & 
  $E$ & 
  $\lbrace P_{123},P_{132} \rbrace$ & 
  $\lbrace P_{12},P_{13},P_{23} \rbrace$ \\
  \hline\\[-0.35cm]         
  $A_1$  & 1 & 1     & 1 \\
  $A_2$  & 1 & 1     & $-$1 \\  
  $E$    & 2 & $-$1  & 0 \\
  \hline\hline\\[-0.35cm]
\end{tabular}
\end{table}

We used the following spin-1/2 angular momentum operators, 
\begin{align}
  \hS_{X}
  = 
  \frac{1}{2}
  \left(%
  \begin{matrix} 
    0 & 1 \\
    1 & 0 
  \end{matrix}   
  \right)
  \quad
  \hS_{Y}= 
  \frac{1}{2}
  \left(%
  \begin{matrix} 
        0 & \iim \\
    -\iim & 0 
  \end{matrix}
  \right)
  \quad
  \hS_{Z}
  =
  \frac{1}{2}
  \left(%
  \begin{matrix} 
    -1 & 0 \\
     0 & 1
  \end{matrix}
  \right) \; .
\end{align}

\section{A simple analytic representation of the electric dipole polarizability matrix}
\noindent
The following functional form was used to fit the electric dipole polarizability computed with electronic structure theory and rotated to the bisector embedding used in the rovibrational part. The $\alpha_{ij}\ (i,j=x,y,z)$ matrix is symmetric, and $\alpha_{xy}=\alpha_{xz}=0$ in our bisector embedding.
The following functions were used for $\alpha_{xx}$, $\alpha_{yy}$  and $\alpha_{zz}$, ($i=x,y,z$)
\begin{align}
  &\alpha_{ii}(\rone,\rtwo,\tangle)
  \nonumber\\
  &\quad 
  =
  \sum_{n_{\rone}=0}^{14}\sum_{n_{\rtwo}=0}^{14}\sum_{n_{\tangle}=0}^{17} H^{(ii)}_{n_{\rone},n_{\rone},n_{\tangle}}
  (\rone-2.5)^{n_{\rone}}
  (\rtwo-2.5)^{n_{\rtwo}}
  (\tangle-60)^{n_{{\tangle}}} \; ,
\end{align}
and for the non-zero off-diagonal element, 
\begin{align}
  &\alpha_{yz}(\rone,\rtwo,\tangle)
  \nonumber\\
  &\quad 
  =
  \sum_{n_{\rone}=0}^{14}\sum_{n_{\rtwo}\ne n_{\rone}=0}^{14}\sum_{n_{\tangle}=0}^{17} 
  H^{(yz)}_{n_{\rone},n_{\rone},n_{\tangle}}
  (\rone-2.5)^{n_{\rone}}
  (\rtwo-2.5)^{n_{\rtwo}}
 (\tangle-60)^{n_{{\tangle}}} \; .
\end{align}

\section{Pauli-allowed states: Calculation of the effect of the $P_{23}$ permutation operator}
In this work, we used PO-DVR for $\rone$ and $\rtwo$ in the (ro)vibrational computations, 
and we need to carry out the symmetry analysis over the quadrature grid.
The proton-spin-rovibrational wave functions (under the effect of an external magnetic field) transform according to the irreps of the $S_{3}$ permutation group. Only the states of $A_{2}$ symmetry fulfil the Pauli principle, \emph{i.e.,} anti-symmetry upon the exchange of any pairs of protons.
$A_{2}$ is a one-dimensional irrep, and therefore, all the rovibrational-spin wave functions allowed by the Pauli principle are singly degenerate (though accidental degeneracies may occur). 
So, we first identify the singly degenerate rovibrational-hyperfine-Zeeman states, and then, check whether they change sign upon the action of one (any) of the particle exchange operations. In the present coordinate and grid representation, the effect of the $P_{23}$ permutation can be calculated, which is outlined in the following paragraphs.

The  rovibrational-spin wave functions are expanded as the linear combination of the product of functions, 
\begin{align}
  \Phi_{J,n_{J},\wK,\tau}(\rone,\rtwo,\cos\tangle) |J,\wK,\tau,M\rangle_\Omega |\Sf_{i}\rangle \; .
  \label{eq:factors}
\end{align}
The effect of $P_{23}$ can be straightforwardly calculated if all factors are eigenfunctions of $P_{23}$. In the case of the one-dimensional irreps, the effect of $P_{23}$ is
\begin{align}
  P_{23}\Phi_{J,n_{J},\wK,\tau}(\rone,\rtwo,\cos\tangle) 
  &=
  \pm\Phi_{J,n_{J},\wK,\tau}(\rone,\rtwo,\cos\tangle) \\
  P_{23} |J,\wK,\tau,M\rangle 
  &=
  \pm |J,\wK,\tau,M\rangle \\
  P_{23} |\Sf_{i}\rangle &= 
  \pm |\Sf_{i}\rangle \; ,
\end{align}
where we used the linear combination of the eight spin functions that are eigenfunctions of $P_{23}$, 
\begin{align}
|\Sf_{1} \rangle&=|\alpha,\alpha,\alpha \rangle \nonumber \\
|\Sf_{2} \rangle&=\frac{1}{\sqrt{2}}(|\alpha,\alpha,\beta\rangle + |\alpha,\beta,\alpha \rangle)  \nonumber \\
|\Sf_{2} \rangle&=\frac{1}{\sqrt{2}}(|\alpha,\alpha,\beta\rangle - |\alpha,\beta,\alpha \rangle)  \nonumber \\
|\Sf_{4} \rangle&=|\alpha,\beta,\beta \rangle \nonumber \\
|\Sf_{5} \rangle&=|\beta,\alpha,\alpha \rangle \nonumber \\
|\Sf_{6} \rangle&=\frac{1}{\sqrt{2}}(|\beta,\alpha,\beta \rangle + |\beta,\beta,\alpha \rangle )\nonumber \\
|\Sf_{7} \rangle&=\frac{1}{\sqrt{2}}(|\beta,\alpha,\beta \rangle - |\beta,\beta,\alpha \rangle )\nonumber \\
|\Sf_{8} \rangle&=|\beta,\beta,\beta \rangle \; .
\end{align}

Regarding degenerate pairs (labelled with $a$ and $b$) of rovibrational states, the $(\wK\tau)$ vibrational blocks, 
$\Phi_{J,n_{J,a},\wK,\tau}(\rone,\rtwo,\cos\tangle)$ and $\Phi_{J,n_{J,b},\wK,\tau}(\rone,\rtwo,\cos\tangle)$ are not necessarily symmetric or antisymmetric to $P_{23}$. But since the states are degenerate, we are free to choose any linear combination of them. So we choose the linear combination for which the matrix representation of $P_{23}$ is diagonal,
\begin{align}
|J,n_{J,1},M \rangle =\cos\phi |J,n_{J,a},M \rangle + \sin\phi |J,n_{J,b},M \rangle 
\end{align}
and
\begin{align}
|J,n_{J,2},M \rangle =-\sin\phi |J,n_{J,a},M \rangle + \cos\phi |J,m_{J,b},M \rangle 
\end{align}
with 
\begin{align}
P_{23}|J,n_{J,1},M \rangle = |J,n_{J,1},M \rangle\nonumber \\
P_{23}|J,n_{J,2},M \rangle =-|J,n_{J,2},M \rangle \; .
\end{align}

Regarding the rotational basis functions, 
the effect of $P_{23}$ on the bisector embedding (Fig.~1) used in this work is the change of $\gamma$ to $\gamma+\pi$, which translates to 
\begin{align}
  P_{23}|J,\tK,M\rangle_\Omega = (-1)^{\tK}|J,\tK,M\rangle_\Omega
\end{align}
and 
\begin{align}
  P_{23}|J,\wK,\tau,M\rangle_\Omega = (-1)^{\wK}|J,\wK,\tau,M\rangle_\Omega \; .
\end{align}

With these considerations, the effect of $P_{23}$ can be determined for each factor in the basis set expansion, Eq.~\eqref{eq:factors}, and the states can be assigned to $A_{1}$ or $A_{2}$ symmetry without ambiguity.  The only ambiguity remains with the identification of the singly degenerate proton-spin-rovibrational states (in an external magnetic field), since accidental degeneracies may occur. In future work, we will implement the full $S_3$ projector, which will provide a fully automated and robust selection of the Pauli allowed subspace, by using the Pekeris coordinates \cite{Pe58,WeCa94} or transformation to FBR (of another DVR computation).


%% file: hyperfine_h3p.bbl
\begin{thebibliography}{70}
\bibitem [{som()}]{som}%
  \BibitemOpen
  \href@noop {} {\bibinfo {title} {See {S}upplemental {M}aterial for further
  implementation and computational details; {t}he full list of computed
  hyperfine-magnetic energies, assignments and {R}aman transition moments are
  deposited in plain text files.}}\BibitemShut {Stop}%
\end{thebibliography}
